# Optimisation de la taille de la série: illustration par un cas industriel de sous-traitance mécanique

**Lyonnet B.[1], Pillet M.[1], Pralus M.[1]**

[1]*Université de Savoie, 74940 Annecy-le-Vieux, France*
barbara.lyonnet@univ-savoie.fr
maurice.pillet@univ-savoie.fr
magali.pralus@univ-savoie.fr

*Résumé:* - Réduire les coûts de revient des produits fabriqués est une des problématiques essentielles des entreprises d'aujourd'hui. Les entreprises de décolletage (entreprises de sous-traitance mécanique) sont confrontées au dilemme suivant : produire juste la demande client ou produire plus. Instinctivement ces entreprises, dont les temps de changement de série sont élevés, cherchent à produire plus que la demande exigée par le client. Cette stratégie leur permet de répondre à des demandes prévisionnelles et réduire ainsi le coût de revient des produits. Ces entreprises réalisent un pari sur les opportunités de vente et pensent réaliser un gain supplémentaire en réalisant des stocks. Nous avons cherché dans cette étude à fournir des éléments de décision pour connaître les limites de cette règle de gestion. Notre proposition porte sur le développement d'un modèle d'aide à la décision prenant en considération le mixte entre opportunités commerciales, contraintes économiques et contraintes de moyen. Ce modèle souligne l'importance particulière du taux de possession et du risque de non vente.

*Abstract:* - Reducing costs of manufactured products is one of the key issues of companies. Bar turning companies (mechanical subcontracting companies) are faced with the following dilemma: use a pull strategy or use a push strategy. Instinctively these companies produce more than demand required by customers. This strategy allows them to respond to requests forecasts and reduce their cost of changeover time. These companies make a bet on sales opportunities and think to realize an additional profit. We have tried to find in this study to provide elements to know the limits of this strategy. Our proposal focuses on developing a model to support the decision taking into account the mix of opportunities, economic constraints and mean constraints. This model features the particular importance of high rates of ownership and the risk of not selling.

*Mots clefs:* **taille de la série, flux tiré, flux poussé, contraintes économiques, opportunités commerciales**

## 1 INTRODUCTION

Un des objectifs majeurs des entreprises d'aujourd'hui est la réduction des coûts de revient des produits fabriqués. Actuellement, la fabrication de ces produits est réalisée selon deux grandes approches : la production à flux tirés et la production à flux poussés.

La production à flux poussés repose sur le déclenchement de la production avant la demande du client. C'est le cas de nombreuses entreprises, notamment dans le secteur du décolletage (entreprises de sous-traitance mécanique) qui fabriquent des produits nécessitant un temps de changement de série élevé et pour lesquelles il est donc intéressant d'anticiper la demande du client. Les pièces produites pourront être stockées et vendues plus tard sur opportunités commerciales. Cette méthode de production peut présenter l'avantage de réduire les coûts de revient des produits fabriqués mais à l'inconvénient d'augmenter les coûts d'immobilisation du stock (Babai, 2008). Cette augmentation peut dans certains cas conduire à l'effet inverse et alors augmenter les coûts de revient.

La production à flux tirés est, quant-à-elle, basée sur la demande réelle des clients, réduisant de ce fait les coûts induits par les stocks excédentaires de produits finis (Ohno, 1988 ; Womack, 1996 ; Shah et Ward, 2007). Cette politique permet ainsi de se dégager des coûts d'immobilisation de stocks excédentaires. Les entreprises de sous-traitance mécanique sont particulièrement confrontées au choix suivant : faut-il produire au plus juste en fonction de la demande réelle ou produire plus et stocker ? A l'heure actuelle la production au plus juste est préconisée par une démarche largement déployée dans les entreprises : la démarche Lean (Womack, 1996). Le cœur de cette démarche repose sur l'élimination des gaspillages dont ceux engendrés par les stocks excédentaires. La production à flux poussés permet quant-à-elle de répondre aux opportunités commerciales et ainsi de réduire les coûts de changement de série. Une telle approche peut permettre aux entreprises de bénéficier d'un avantage concurrentiel. Toutefois, cette stratégie présente un risque non négligeable si l'entreprise ne revend pas rapidement l'ensemble de son stock excédentaire ; elle s'expose à des pertes financières. L'orientation vers une de ces deux approches de production reste délicate. Ces approches sont rarement utilisées d'une manière exclusive (Zhao et al., 2005). Plusieurs études ont montré qu'il fallait intégrer ces deux méthodes de gestion (Siala et al., 2006 ; Ball et al., 2004). Le compromis à réaliser entre ces approches de production doit prendre en considération les coûts de changement de série et les coûts de stockage. Pour trouver le meilleur équilibre entre ces deux éléments, une méthode connue sous le nom de la quantité économique à commander ou formule de Wilson a été développée (Erlenkotter, 1989 ; Diallo, 2006). Ce modèle conduit à un calcul de taille de lot (Giard, 2006). Initialement prévu pour calculer une quantité à commander, il peut être appliqué pour identifier une quantité à produire (Javel, 2003). Cependant, cette méthode est inadéquate lorsque la demande est variable au cours du temps (Wagner et Whitin, 1958 ; Silver et Meal, 1973). De la même manière, les coûts et gains liés aux opportunités de ventes sont ignorés (Babai, 2008). Pour ces raisons il apparaît nécessaire de développer une nouvelle approche permettant d'évaluer la quantité optimale à fabriquer au moment d'une commande client.

Notre étude consiste à répondre aux questions suivantes : Y a-t-il encore une place pour la règle de gestion d'anticipation de commandes potentielles futures? L'approche Lean qui suggère de ne produire que la quantité nécessaire doit-elle être appliquée quelle que soit le produit, sa régularité, son histoire? Si place il y a pour la règle de gestion intuitivement suivie par les décolleteurs, peut-on formaliser la démarche pour fournir une aide à la décision? Pour répondre à ces questions, il faut prendre en considération les coûts de stockage, les coûts de changement de série, les demandes et opportunités de vente et les risques associés à la surproduction. Aussi, les contraintes de moyens matériels ne doivent pas être ignorées. En effet, aucune fabrication n'est possible sans ces dernières.

Pour ce faire, nous nous proposons de développer un modèle de décision assimilant le mixte entre opportunités et contraintes. Les éléments de réponse à ces questions conduisent à l'élaboration d'un outil d'aide à la décision, permettant de choisir la stratégie de production la plus adaptée.

**2 METHODES**

La démarche de recherche engagée dans cette étude est fondée sur l'observation de situations réelles dans le secteur du décolletage (entreprise de sous-traitance mécanique). Pour généraliser ces observations, nous avons mis au point un modèle de calcul de la quantité optimale à produire à partir des contraintes économiques, des opportunités de vente, et des contraintes de moyens.

*2.1 Principales caractéristiques de l'entreprise étudiée*

L'entreprise de décolletage étudiée, située au cœur de la vallée de l'Arve (Haute-Savoie), constitue un terrain d'application particulièrement intéressant. La vallée de l'Arve est considérée comme l'un des principaux systèmes productifs locaux français. Les entreprises de la Vallée de l'Arve fournissent plus de 60% du chiffre d'affaires français de l'activité du décolletage, c'est à dire l'usinage de pièces mécaniques à partir de barres essentiellement métalliques.

*2.2 Hypothèses du modèle*

Un préalable à cette étude a été de décrire le cadre de fonctionnement courant de l'entreprise étudiée. Les hypothèses suivantes sont prises en compte (hypothèses validées sur le terrain) :
- les temps de changement de série sont longs > 4h et difficilement compressibles
- les produits supplémentaires à la quantité commandée sont stockés avant d'être vendus
- les quantités commandées sont considérées livrées immédiatement après fabrication et ne subissent pas de coût de stockage

*2.3 Contraintes économiques*

Les contraintes économiques prises en compte par notre modèle sont représentées par le calcul du coût de revient des produits fabriqués.
Le coût de revient est obtenu par l'addition de trois autres coûts :
Coût de revient total= Coût d'achat des matières premières
+ Coût de production
+ Coût hors production
Où :
- les coûts d'achat des matières premières correspondent au prix facturé par le fournisseur augmenté des frais spécifiques liés à l'achat hors frais de structure (frais de transports, assurance, frais de douane…)
- les coûts de production correspondent à la valeur ajoutée (main d'œuvre, coût machine dont coûts de changement de série, plus les frais généraux spécifiques à la production)
- les coûts hors production du produit correspondent aux frais généraux hors production tels que les frais de distribution ou d'administration (Javel, 2003).
Dans le secteur du décolletage, les temps de changement de série étant longs et difficilement compressibles, nous nous sommes concentrés sur l'impact du coût de changement de série sur le coût de revient. Ce coût est indépendant de la quantité de pièces produites, il dépend principalement du temps nécessaire au changement de série.

Dans le cas d'une production à flux tirés, on établit que le coût de revient (CR) par pièce est fonction de la quantité commandée (QC) et du coût de changement de série (CS):

$$CR = \frac{CS + CU*QC}{QC} \quad \text{(Eq 1)}$$

Avec :

$CU$ = Coût de revient unitaire hors coût de changement de série.

On en déduit la relation :

$$CU = CR - \frac{CS}{QC} \quad \text{(Eq 2)}$$

Dans le cas d'une production à flux poussés, les pièces supplémentaires sont stockées avant d'être vendues, un coût de possession doit alors être calculé.

Le coût total de possession (CP) des pièces supplémentaires (X) pour la période s'établit en fonction du taux de possession (i) et du coût de revient hors coût de changement de série (CU) :

$$CP = (CU*X)*i \quad \text{(Eq 3)}$$

Dans cette relation le coût de changement de série n'est pas pris en considération. En effet, dans le fonctionnement réel de l'entreprise étudiée le coût de changement de série est vendu avec les pièces commandées. Le taux de possession annuel « i% » est le coût de possession ramené à un euro de produits stockés. Il est obtenu en divisant le coût total des frais de possession par le stock moyen. Ces frais couvrent :
- l'intérêt du capital immobilisé
- les coûts de magasinage (loyer et entretien des locaux, assurance, frais de personnel et manutention)
- les détériorations du matériel
- les risques d'obsolescence

Le taux utilisé dans les entreprises se situe entre 15% et 35% suivant le type de pièce et la qualité de la gestion des stocks (Javel, 2003).

Dans ce contexte, on établit que le coût de revient par pièce est fonction de la quantité supplémentaire (X) et du coût de possession (CP) affecté :

$$CR = \frac{CS + CU*(QC+X)}{QC+X} + \frac{CP}{QC+X} \quad \text{(Eq 4)}$$

D'où,

$$CR = \frac{CS + CU*(QC+X)}{QC+X} + \frac{(CU*X)*i}{QC+X} \quad \text{(Eq 5)}$$

Par suite,

$$CR = \frac{CU*X(i+1) + QC*CU + CS}{QC+X} \quad \text{(Eq 6)}$$

Notre modèle vise à rechercher la quantité X de pièces supplémentaires à produire pour laquelle le coût de revient est minimal.

C'est à dire la valeur de X pour laquelle la dérivée du coût de revient par rapport à la quantité X est nulle.

D'où,

$$\frac{dCR}{dX} = \frac{i\,Cu\,Qc - Cs}{(X+Qc)2} = 0 \quad \text{(Eq 7)}$$

Dans cette équation, X étant au dénominateur, cette dérivée s'annule lorsque X est infini.

Par contre le dénominateur étant strictement positif, le signe de la dérivée dépend du numérateur. On souhaite avoir une dérivée négative, correspondant à une diminution du coût de revient en fonction de X.

Ce qui conduit à la relation suivante qui exprime la condition pour qu'il soit intéressant d'envisager de produire sur stock :

$$i \leq \frac{CS}{QC.Cu} \quad \text{(Eq 8)}$$

Lorsque cette condition est obtenue l'entreprise réalise un gain financier malgré le coût de stockage. L'équation 8 ainsi calculée est indépendante de la quantité X fabriquée en plus. En effet, la quantité de pièces supplémentaires fabriquées dépend du taux de possession des stocks (i) par l'entreprise.

La stratégie qui consiste à produire plus et stocker engendre cependant un risque supplémentaire de perte financière liée au risque de non vente. Pour prendre en considération ce risque nous proposons d'identifier les gains et pertes associés à un risque de non vente estimée. Retrouver le meilleur compromis en prenant considération le risque de non vente consiste à évaluer deux cas : le cas où les pièces fabriquées en plus se vendent et le cas où l'entreprise ne peut les vendre.

Dans le cas où les pièces fabriquées en plus (X) se vendent l'entreprise peut réaliser soit un gain soit une perte financière en fonction du taux de possession (Cf. Eq 8). Ce résultat financier (R) réalisé par l'entreprise dépend du coût de changement de série (CS) et du coût total de possession du stock (CP). En effet, en produisant plus que la quantité commandée, l'entreprise économise le coût de changement de série (vendue lors de la première commande) mais perd de l'argent en fonction du coût total de possession. On en déduit la relation :

$$R = CS - CP \quad \text{(Eq 9)}$$

Avec,

$$CP = (CU*X)*i \quad \text{(Eq 3)}$$

En prenant en considération la probabilité de vente (P) on peut établir:

$$R' = (CS - (CU*X)*i))*(P) \quad \text{(Eq 10)}$$

Dans le cas où les pièces supplémentaires (X) ne se vendent pas l'entreprise réalise une perte financière (PR) pouvant être exprimée comme suit :

$$PR = (CU*X(1+i))*(1-P) \quad \text{(Eq 11)}$$

Avec,

*1-P= probabilité de non vente*

Pour prendre en considération ces deux relations, on peut établir que l'entreprise réalise un gain lorsque les gains (G) financiers potentiels sont plus importants que les pertes. On en déduit la relation suivante :

$$G = R' - PR \quad \text{(Eq 12)}$$

*2.4 Contraintes de moyens*

Un des principes de base d'une bonne gestion réside dans le fait de ne pas prendre du temps sur une machine pour faire de la vente virtuelle alors que l'on peut faire une vente réelle. Pour prendre en considération ces contraintes de moyens, notre modèle s'appuie sur le temps disponible (TD) des ressources nécessaires à la fabrication des pièces et le temps de cycle de la pièce (TC). Ce temps disponible est établi à partir de la ressource identifiée comme goulot. En effet, la ressource goulot limite l'ensemble du flux, si celle-ci n'est pas disponible, le flux de fabrication de la pièce est arrêté (Goldratt, 2006; Marris, 2005). Le temps disponible doit donc prendre en compte le taux d'occupation prévisionnel de la ressource goulot.

Notre modèle cherche la quantité de pièces supplémentaires pouvant être produites selon le temps disponible des ressources.

Pour ce faire, l'équation suivante peut donc être proposée :

$$X \leq \frac{TD}{TC} \qquad (Eq\ 13)$$

*2.5 Prise en considération du mixte entre contraintes et opportunités*

L'objectif de notre modèle de décision est la prise en considération des deux éléments précédemment présentés :
- contraintes économiques et opportunités de vente
- contraintes de moyens

Pour maximiser les gains financiers des entreprises, la quantité optimale de pièces à produire doit prendre en compte le mixte de contraintes et d'opportunités dont les équations sont les suivantes :

$$G = R' - PR \qquad (Eq\ 12)$$
$$X \leq \frac{TD}{TC} \qquad (Eq\ 13)$$

## 3 RESULTATS

*3.1 Données recueillies*

Pour illustrer notre modèle d'aide à la décision sur la quantité à produire, nous avons réalisé deux cas d'application à partir des données de l'entreprise de décolletage étudiée. Les données recueillies pour la fabrication des pièces nommée *a* et *b* sont présentées dans le tableau 1. Dans notre cas la pièce fabriquée *a* possède un coût de changement de série de 270€ et la pièce *b* de 2000€. Le coût de revient unitaire de chaque pièce hors coût de changement de série est égal à 0,06€ alors que celui de la pièce *b* est de 0,3€. Dans les deux cas présentés, l'entreprise a reçu une commande de 20000 pièces qui sera livrée immédiatement après la fabrication.

**Tableau 1. Données recueillies pour la fabrication de la pièce *a***

| Pièces | | a | b |
|---|---|---|---|
| Coût de changement de série | CS | 270 € | 2 000 € |
| Coût de revient hors coût de changement de série par pièce | CU | 0,06 € | 0,3 |
| Quantité commandée en nombre de pièces | QC | 20000 | 20000 |
| Coût de revient (dans le cas d'une fabrication au plus juste) | CR | 0,074 € | 0,400 € |
| Taux de possession affecté pour un an d'immobilisation | i% par an | 9% | 9% |

Le coût de revient calculé par l'entreprise pour cette commande est de 0,07€ pour la pièce *a* et pour la pièce *b* de 0.4€. Ce coût de revient est obtenu lorsque l'entreprise produit en fonction de la demande réelle exigée par le client.

*3.2 Résultats obtenus à partir des contraintes économiques et opportunités de ventes*

Dans le cas où l'entreprise produirait des pièces supplémentaires le taux de possession affecté pour un an d'immobilisation est de 9% pour les deux pièces. Ce taux permet de calculer le coût de stockage des pièces fabriquées en plus de la quantité commandée.

A partir de ces données, le coût de revient par pièce en cas de fabrication supplémentaire à la demande du client est calculé selon l'équation 6 (Cf. Tableaux 2 et 3). Dans les deux cas présentés, malgré l'ajout d'un coût de stockage, le coût de revient des pièces *a* et *b* est inférieur à celui obtenu lorsque l'entreprise fabrique au plus juste. Pour la pièce *a* le coût de revient des pièces produites en plus de la demande client est de 0.068€ et de 0.333€ pour la pièce *b*. Ce résultat est en accord avec le risque lié à l'immobilisation calculé avec l'équation 8.

D'après les calculs présentés dans le tableau 2, l'entreprise réaliserait un gain financier lorsque la probabilité de vente de la pièce *a* est égale ou supérieure à 90% pour une durée d'immobilisation de 150 jours. Pour la pièce *b* l'entreprise réaliserait un gain financier à partir d'une probabilité de vente de 80%. Pour les deux pièces, quelque soit la probabilité de vente les coûts de stockage n'engendrent pas de perte financière. En effet, les résultats financiers en cas de vente sont toujours positifs malgré un taux de possession du stock correspondant à 3,7%.

Les gains financiers potentiels pour la fabrication de pièces *a* sont d'environ 78€ lorsque la probabilité de vente est de 90% et d'environ 225€ lorsque la probabilité de vente est de 100%. Pour la pièce *b* les gains financiers sont de 178€ environ pour une probabilité de vente de 80%, de 978€ pour une probabilité de vente de 90% et de 1778€ pour une probabilité de vente de 100%.

**Tableau 2. Résultats obtenus pour la pièce a**

| | | | | | | | | | | |
|---|---|---|---|---|---|---|---|---|---|---|
| Durée de stockage maximale | 150 | 150 | 150 | 150 | 150 | 150 | 150 | 150 | 150 | 150 |
| Quantité supplémentaire (X) | 20000 | 20000 | 20000 | 20000 | 20000 | 20000 | 20000 | 20000 | 20000 | 20000 |
| Probabilité de ventes (P) | 10% | 20% | 30% | 40% | 50% | 60% | 70% | 80% | 90% | 100% |
| Nombre total de pièces à fabriquer | 40000 | 60000 | 80000 | 100000 | 120000 | 140000 | 160000 | 180000 | 200000 | 220000 |
| Taux de possession pour la période de stockage (i%) | 3,7% | 3,7% | 3,7% | 3,7% | 3,7% | 3,7% | 3,7% | 3,7% | 3,7% | 3,7% |
| Coût de revient de la pièce (CR calculé par l'équation 6) | 0,068 € | 0,068 € | 0,068 € | 0,068 € | 0,068 € | 0,068 € | 0,068 € | 0,068 € | 0,068 € | 0,068 € |
| Risque lié à l'immobilisation maximale (calculé par l'équation 8) | 0,225 € | 0,225 € | 0,225 € | 0,225 € | 0,225 € | 0,225 € | 0,225 € | 0,225 € | 0,225 € | 0,225 € |
| Résultat financier en cas de vente (R' calculé par l'équation 10) | 22,56 € | 45,12 € | 67,68 € | 90,25 € | 112,81 € | 135,37 € | 157,93 € | 180,49 € | 203,05 € | 225,62 € |
| Perte financière en cas de non vente (PR calculé par l'équation 11) | 1 119,95 € | 995,51 € | 871,07 € | 746,63 € | 622,19 € | 497,75 € | 373,32 € | 248,88 € | 124,44 € | - € |
| Gain si on fabrique par anticipation (G calculé par l'équation | - 1 097,38 € | - 950,38 € | - 803,38 € | - 656,38 € | - 509,38 € | - 362,38 € | - 215,38 € | - 68,38 € | 78,62 € | 225,62 € |

**Tableau 3. Résultats obtenus pour la pièce b**

| | | | | | | | | | | |
|---|---|---|---|---|---|---|---|---|---|---|
| Durée de stockage maximale | 150 | 150 | 150 | 150 | 150 | 150 | 150 | 150 | 150 | 150 |
| Quantité supplémentaire (X) | 20000 | 20000 | 20000 | 20000 | 20000 | 20000 | 20000 | 20000 | 20000 | 20000 |
| Probabilité de ventes (P) | 10% | 20% | 30% | 40% | 50% | 60% | 70% | 80% | 90% | 100% |
| Nombre total de pièces à fabriquer | 40000 | 60000 | 80000 | 100000 | 120000 | 140000 | 160000 | 180000 | 200000 | 220000 |
| Taux de possession pour la période de stockage (i%) | 3,7% | 3,7% | 3,7% | 3,7% | 3,7% | 3,7% | 3,7% | 3,7% | 3,7% | 3,7% |
| Coût de revient de la pièce (CR calculé par l'équation 6) | 0,333 € | 0,333 € | 0,333 € | 0,333 € | 0,333 € | 0,333 € | 0,333 € | 0,333 € | 0,333 € | 0,333 € |
| Risque lié à l'immobilisation maximale (calculé par l'équation 8) | 35,55% | 35,55% | 35,55% | 35,55% | 35,55% | 35,55% | 35,55% | 35,55% | 35,55% | 35,55% |
| Résultat financier en cas de vente (R' calculé par l'équation 10) | 177,81 € | 355,62 € | 533,42 € | 711,23 € | 889,04 € | 1 066,85 € | 1 244,66 € | 1 422,47 € | 1 600,27 € | 1 778,08 € |
| Perte financière en cas de non vente (PR calculé par l'équation 11) | 5 599,73 € | 4 977,53 € | 4 355,34 € | 3 733,15 € | 3 110,96 € | 2 488,77 € | 1 866,58 € | 1 244,38 € | 622,19 € | - € |
| Gain si on fabrique par anticipation (G calculé par l'équation | - 5 421,92 € | - 4 621,92 € | - 3 821,92 € | - 3 021,92 € | - 2 221,92 € | - 1 421,92 € | - 621,92 € | 178,08 € | 978,08 € | 1 778,08 € |

*3.3 Résultats obtenus à partir des contraintes de moyens matériels*

Les données prises en considération concernant les contraintes de moyens matériels sont résumées dans le Tableau 4.

**Tableau 4. Données recueillies pour la prise en compte des contraintes de moyens**

| Données nécessaires | | Pièce a | Pièce b |
|---|---|---|---|
| Quantité commandée en pièces | QC | 20000 | 20000 |
| Temps de cycle de la pièces en minute | TC | 0,3 | 0,5 |
| Temps disponible en minute | TD | 2000 | 12000 |

D'après l'équation 13 les résultats pour la pièce *a* sont les suivants :

$$\frac{TD}{TC} = \frac{2000}{0,3}$$

$$\frac{TD}{TC} = 6666$$

Soit,

$X \leq 6666$ pièces

Pour la pièce *b* les résultats sont les suivants :

$$\frac{TD}{TC} = \frac{12000}{0,5}$$

$$\frac{TD}{TC} = 24000$$

Soit,

$X \leq 24000$ pièces

Les contraintes de moyens matériels limitent donc la production de pièces *a* supplémentaires à 6666 et de pièces *b* à 24000.

A partir de ce mixte entre opportunités et contraintes l'entreprise peut désormais décider de la stratégie la plus favorable. La prise en compte des contraintes économiques nous a permis d'identifier une quantité de pièces supplémentaires (X) pouvant être fabriquées en fonction des opportunités commerciales pour une probabilité de vente supérieure ou égale à 90% pour la pièce *a* et une probabilité de vente supérieure ou égale à 80% pour la pièce *b*. Dans notre exemple le coût de stockage des pièces fabriquées en plus génèrent peu de perte financière sur une durée d'immobilisation inférieure ou égale à 150 jours. La seconde étape de notre modèle a été la prise en compte des contraintes de moyens matériels. Dans notre cas d'application, l'entreprise ne peut pas fabriquer plus de 6666 pièces *a* et 24000 pièces *b* supplémentaires à la demande. Dans cet exemple l'entreprise est limitée par ses contraintes de moyens matériels pour la pièce a. Comme l'entreprise fabrique par lot de 20000, elle devra se limiter à la production de pièces *a* en fonction des demandes clients. En revanche, pour la pièce *b* l'entreprise pourra produire 20000 pièces supplémentaires à la demande si la probabilité de vente sur une durée de 150 jours maximum est supérieure ou égale à 80%.

## 4 DISCUSSION

Plusieurs raisons nous ont conduits au développement de notre modèle. Jusqu'à présent les entreprises de décolletage réalisaient un pari sur le gain potentiel généré par leur production supplémentaire qui repose sur leurs ventes potentielles. Instinctivement, ces entreprises cherchent à produire plus que la demande exigée par le client. Pourtant, les concepts tels que ceux employés dans la démarche Lean manufacturing incitent à ne produire qu'en fonction de la demande du client. En effet, la surproduction, de même que les stocks excédentaires sont considérés dans cette démarche comme source de gaspillages (Ohno, 1988 ; Womack, 1996; Drew et al., 2004 ; Hicks, 2007). L'ensemble de cette profession se tromperait-elle dans son mode de gestion ? Nous avons cherché dans cette étude à fournir des éléments de décision pour connaître les limites de cette règle de gestion.

Nous avons ainsi développé une nouvelle démarche permettant d'optimiser les coûts de revient en prenant en compte un mixte entre opportunités et contraintes pour calculer une quantité optimale à produire. Ce modèle souligne l'importance du taux de possession et plus particulièrement de la probabilité de vente (P). Grâce à cette méthode, une entreprise dont les coûts de changement de série sont élevés peut connaître le meilleur compromis entre les deux approches de production : flux tiré et flux poussé. Ce compromis est réalisé en prenant en considération les contraintes économiques telles que les coûts de changement de série, les coûts de stockage, les demandes clients et les opportunités commerciales de vente. Un grand avantage de notre modèle d'aide à la décision est la prise en considération des opportunités commerciales. Dans notre étude, cette estimation repose sur la prise en compte de la probabilité de non vente et conduit à une estimation de l'impact des paris réalisés par les entreprises.

Notre modèle est parfaitement adapté aux entreprises disposant d'un coût de changement de série élevé. En revanche, pour les entreprises disposant de coût de changement de série faible, le modèle présente moins d'intérêt, l'évolution du coût de revient des produits en cas de fabrication supplémentaire à la demande est prévisible. Pour ces dernières, une fabrication au plus juste est recommandée. De plus, dans notre cas d'application, même avec un coût de changement de série important, l'entreprise de décolletage étudiée réalise un gain financier lorsque le risque de non vente est inférieur à 20%. L'entreprise doit donc être sûre à 80% des estimations réalisées pour s'engager dans cette stratégie ; le cas contraire la conduirait à une perte. Cette très faible marge d'incertitude possible sur les ventes futures, même avec un coût de changement de série important montre bien l'intérêt d'appliquer les principes du Lean manufacturing. Il existe cependant une zone correspondant à des situations bien connues des décolleteurs justifiant les pratiques de gestion appliquées empiriquement. Malheureusement dans les études réalisées, l'empirisme conduit souvent à un sur stockage excessif, l'approche proposée permettrait d'apporter un compromis plus raisonnable.

## 5. CONCLUSION

Cette approche de formalisation du problème réel de production rencontré dans le cas de temps de changement de séries coûteux, nous a conduits à décrire les règles de décisions adaptées. Ces règles de décision prennent en considération un mixte entre opportunité de ventes, contraintes économiques et contraintes de moyens matériels. Il est très facile, à partir des équations démontrées dans ce papier, de créer un outil d'aide à la décision à l'aide d'un simple tableur. Il nous semble recommandable d'inciter à l'application des règles de gestion développées pour les entreprises disposant de coûts de changement de série important.

## 6 REMERCIEMENTS

## 7 REFERENCES

Babai M. Z., (2008) Politique de pilotage de flux dans les chaines logistiques : impact de l'utilisation des prévisions sur la gestion des stocks, *thèse de doctorat*.

Ball M.O., Chen C.-Y. and Zhao Z.-Y., (2004) Available to Promise. Handbook of Supply Chain Analysis in the eBusiness Era. In D. Simchi-Levi, S. David Wu, and M. Shen (eds). *Boston: Kluwer Academic Publishers*, pp. 447–484.

Diallo C. (2006), développement d'un modèle de gestion des pièces de rechange, *thèse de doctorat*.

Drew J., Mc Callum B., Rogenhoffer S. (2004) Objectif Lean, *Mc Kinsey et company*.

Erlenkotter D. (1989) An early classic misplaced: Ford W. Harris's economie order quantity model of 1915, *Management Science*, vol. 35, pp. 898-900.

Giard V., Mendy G. (2006) Amélioration de la synchronisation de la production sur une chaîne logistique, *Revue française de génie industrielle*.

Goldratt E., (2006) Le but, un processus de progrès permanent, *troisième édition, AFNOR*.

Hicks B.J. (2007) Lean information management : understanding and eliminating waste, *International Journal Of Information Management*.

Javel G. (2003) Pratique de la gestion industrielle: organisation, méthodes et outils, *Dunod*.

Marris P., (2005) Le Management Par les Contraintes, *Editions d'Organisation*.

Ohno, T., (1988) Toyota Production System: Beyond Large-Scale Production, *Productivity Press*.

Shah R and Ward P T. (2007) Defining and developing measures of lean production, *Journal of Operations Management*, 25 (4).

Siala M., Campagne JP., Ghedira K. (2006) Proposition d'une nouvelle approche pour la gestion du disponible dans les chaînes logistiques. *MOSIM'06, Maroc*.

Silver, E. A. et Peterson, R. (1985) Decision Systems for Inventory Management and Production Planning, *John Wiley and Sons, New York*.

Wagner, H. M. et Whitin, T. M. (1958) Dynamic Version of the Economic Lot Size Model, *Management Science*, vol. 5, pp. 89-96.

Womack J., Jones D. (1996) System Lean : Penser l'entreprise au plus juste, *village mondial, 1ère éd*.

Zhao Z-Y, Ball M.O and Kotake M. (2005) Optimization-Based Available-To-Promise with Multi-Stage Resource Availability. *Annals of Operations Research*. 135, p. 65–85.